\documentstyle[twocolumn,aps,prl,epsf]{revtex}

\begin{document}


\title{Enhanced shot noise in resonant tunneling: theory and experiment}
\author{G. Iannaccone\cite{email}, G. Lombardi, M. Macucci, and B. Pellegrini}
\address{
Dipartimento di Ingegneria dell'Informazione: 
Elettronica, Informatica e Telecomunicazioni \\
Universit\`a degli studi di Pisa, Via Diotisalvi 2, 
I-56126 Pisa, Italy}
\date{\today}
\maketitle

\begin{abstract}   
We show that shot noise in a resonant tunneling diode biased in the negative
differential resistance regions of the I-V characteristic is enhanced
with respect to ``full'' shot noise. We provide experimental
results showing a Fano factor up to 6.6, and show that it is
a dramatic effect caused by electron-electron interaction
through Coulomb force, enhanced by the 
particular shape of the density of states in the well. 
We also present numerical results from the proposed theory, which are
in agreement with the experiment, demonstrating that the model 
accounts for the relevant physics involved in the phenomenon.
\end{abstract}


\narrowtext

Deviations from the purely poissonian shot noise (the so-called
``full'' shot noise) in mesoscopic
devices and resonant tunneling structures have
been the subject of growing interest in the last decade.
\cite{lesovik89,li90,yurkkoch90,buettiker90,buettiker92,chenting92,davihyld92,brown92,iannashot97,liu95,ciambrone95,lombardi97}
The main reason is that noise is a very sensitive probe of
electron-electron interaction, \cite{landauer96} both due
to the Pauli principle and to Coulomb force, and provides 
information not obtainable from DC and AC characterization; furthermore, 
noise depends strongly on the details of device structure, so that
the capability of modeling it in nanoscale devices implies
and requires a deep understanding of the collective transport mechanisms 
of electrons.

Almost all published theoretical and experimental studies 
have focussed on the suppression of shot noise due to negative 
correlation between
current pulses caused by single electrons traversing the device.
Such correlation may be introduced by Pauli exclusion, which limits
the density of electrons in phase space, and/or by Coulomb repulsion,
depending
on the details of the structure and on the dominant transport
mechanism,\cite{chenting92,davihyld92,brown92,iannashot97}
and make the pulse distribution sub-poissonian, leading to suppressed
shot noise.

In particular, for the case of resonant tunneling structures,
several theoretical and experimental studies 
have appeared in the literature,
\cite{li90,yurkkoch90,buettiker90,buettiker92,chenting92,davihyld92,brown92,iannashot97,liu95,ciambrone95,lombardi97}
assessing
that the power spectral density
of the noise current $S$ in such devices may be suppressed down to
half the ``full'' shot noise value $S_{\rm full} = 2 q I$ , i.e., that 
associated to a purely poissonian process.

In this Letter, we propose a theoretical model and show experimental 
evidence of the opposite behavior, that is of enhanced shot noise 
with respect to $S_{\rm full}$, which is to be expected in
resonant tunneling structures biased in the negative 
differential resistance region of the $I$-$V$ characteristic.

We shall show that in such condition Coulomb interaction
and the shape of the 
density of states in the well 
introduce positive correlation
between consecutive current
pulses, leading to a super-poissonian pulse distribution, 
which implies a super-poissonian shot noise.

First, we shall show an intuitive physical picture of the phenomenon,
then we shall express it in terms of a model for transport and noise in 
generic resonant tunneling structures presented elsewhere 
\cite{iannashot97,ianna_unified95}. 
Furthermore, we shall show the experimental results,
exhibiting a noise power spectral density almost 6.6 times greater than 
$S_{\rm full}$, and compare it with the results
provided by a numerical implementation of our model.

As is well known, the typical I-V characteristic of a resonant tunneling diode 
is due to the shape of the density of states in the well,
which consists of a series of narrow peaks in correspondence with the 
longitudinal allowed energies in the well: for the
GaAs/Al$_{0.36}$Ga$_{0.64}$As material system considered here, 
there is a single narrow peak.
In the negative differential resistance region of the 
I-V characteristic, the peak of the density of states is below the conduction
band edge of the cathode: with increasing voltage, the density of states
is moved downward, so that fewer states are available for
tunneling from
the cathode, and the current decreases.

\begin{figure}
\epsfxsize = 8.cm
\epsffile{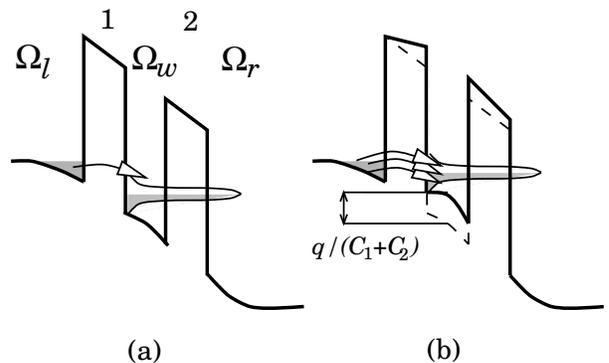}
\vspace{0.2cm}
\caption{Enhanced shot noise is obtained because
an electron tunneling into the well (a) from the 
cathode raises the potential energy of the well by an amount 
$q/(C_1+C_2)$ so that more states are available for tunneling from the
cathode (b)}
\end{figure}

The microscopic mechanism which allows for enhanced shot noise is the
following (see Fig. 1): an electron tunneling into the well from the 
cathode raises the potential energy of the well by an amount 
$q/(C_1+C_2)$, where $q$
is the electron charge, $C_1$ and $C_2$ the capacitances
between the well region and either contacts;
as a consequence, 
the density of states in the well is shifted upwards 
by the same amount, with
the result that more states are available for successive tunneling
events from the
cathode, and the probability per unit time that electrons
enter the well increases. That means that electrons entering 
the well are positively
correlated, so that enhanced shot noise is to be expected.

For a more analytical derivation we can consider the structure 
as consisting of three regions 
$\Omega_l$, $\Omega_w$, and $\Omega_r$, i.e.,
the left reservoir, the well region, and the right reservoir,
respectively, that are only weakly coupled through the two tunneling 
barriers 1 and 2, as sketched in Fig. 1(a).
In addition, we suppose that electron transport is well described
in terms of sequential tunneling (which is reasonable, 
except for the case of temperatures in the millikelvin range): an electron in
$\Omega_l$  traverses barrier 1, loses
phase coherence and relaxes to a quasi-equilibrium energy distribution
in the well region $\Omega_w$, then traverses barrier 2 and 
leaves through $\Omega_r$.

Since confinement is realized only in one direction (that of MBE growth),
a state in $\Omega_s$ ($s=l,r,w$) is characterized by its 
longitudinal energy $E$,  its transverse wave vector ${\bf k_T}$, 
and its spin $\sigma$, and
tunneling can be treated as a transition between
levels in different regions \cite{bardeen61} in which $E$, ${\bf k_T}$ 
and $\sigma$ are conserved.

Following Davies {\em et al.} \cite{davihyld92}, we introduce
``generation'' and ``recombination'' rates through both
barriers: \cite{iannashot97} the generation rate $g_1$ is the transition
rate from $\Omega_l$ to $\Omega_w$, i.e., the sum of the
probabilities per unit time of having a transition from $\Omega_l$
to $\Omega_w$ given by the Fermi ``golden rule'' over all pairs of
occupied states in $\Omega_l$ and
empty states in $\Omega_w$. Analogously, we define $r_1$, 
the recombination rate through barrier 1 (from
$\Omega_w$ to $\Omega_l$), $g_2$ and $r_2$, generation and recombination
rates through barrier 2.

Since negative differential resistance is obtained at
high bias, when the electron flux is one-directional, $r_1$ and
$g_2$ can be discarded, while $g_1$ and $r_2$ are:
\begin{eqnarray}
g_1 & = &  2 \frac{2 \pi}{\hbar} \int  dE \;|M_{1lw}(E)|^2 \rho_l(E)
		\rho_w(E) \times \nonumber \\
	& & 	\int  d{\bf k_T} \rho_T({\bf k_T}) 
		f_l(E,{\bf k_T}) (1 - f_w(E,{\bf k_T})) 
,\label{g1} \\
r_2 & = & 2 \frac{2 \pi}{\hbar} \int dE \;|M_{2rw}(E)|^2 \rho_r(E)
		\rho_w(E) \times \nonumber \\
	& & \int  d{\bf k_T} \rho_T({\bf k_T}) 
		f_w(E,{\bf k_T}) (1 - f_r(E,{\bf k_T}))
.\label{r2}
\end{eqnarray}
where $\rho_s$, $f_s$, ($s=l,w,r$), are the longitudinal
density of states and the equilibrium occupation factor
in $\Omega_s$ (dependent on the quasi Fermi level $E_{fs}$), 
respectively, and 
$\rho_T$ is the density of transverse states;
$ M_{1lw}(E)$ is the matrix 
element for a transition through barrier 1 between 
states of longitudinal energy
$E$: it is obtained in Ref. \cite{ianna_unified95} as
$| M_{1lw}(E)|^2 = \hbar^2 \nu_l(E) \nu_w(E) T_1(E)$, where $\nu_s$ ($s=l,w,r$)
is the so-called attempt frequency in $\Omega_s$ and $T_1$ is the 
tunneling probability
of barrier 1; $M_{2rw}(E)$ is analogously defined.

As is well known, in the negative resistance
region of the $I$-$V$ characteristic, the peak of $\rho_w$ is below the conduction 
band edge of the left electrode.
In such a way, as the voltage is increased, the number of allowed 
states for a transition from
$\Omega_l$ to $\Omega_w$ is reduced, hence the current decreases.
Since all electrons relax to lower energy
states once they are in the well, it is reasonable to
assume that $\Omega_w$-states
with longitudinal energies above the conduction band edge
of the left electrode $E_{\rm cbl}$ are empty, i.e.,
correspond to a zero
occupation factor $f_w$ (analogously, $f_r =0$). 
In addition, if we discard size
effect in the cathode, we have that $2 \pi \hbar \rho_l \nu_l = 1$ 
if $E > E_{\rm cbl}$. Therefore we can rewrite $g_1$ and $r_2$ as
\begin{eqnarray}
g_1 & = & 2 \int_{E_{\rm cbl}}^\infty dE
      \, \nu_w(E) \, \rho_w(E) T_1(E) F_l(E),  
\label{g1a} \\
r_2 & = & 2 \int_{E_{\rm cbw}}^\infty dE
      \, \nu_w(E) \, \rho_w(E) T_2(E) F_w(E)
,\label{r2a}
\end{eqnarray}
where $F_s(E)$ is the occupation factor of $\Omega_s$, ($s=l,w,r$)
integrated over the transverse wave vectors
$F_s(E) \equiv \int d{\bf k_T} \, \rho_T({\bf k_T}) \, f_s(E, {\bf k_T})$, 
and $E_{\rm cbw}$ is the bottom of the conduction band edge in the well.

Let us point out that $g_1$ and $r_2$ depend on the number of
electrons $N$ in the well region both through the potential energy
profile, which is affected by the charge in $\Omega_w$ through
the Poisson equation, and through the term $F_w$ in (\ref{r2a})
which depends on $N$ through the quasi-Fermi level $E_{fw}$.
It is worth noticing that in our case Pauli exclusion has no
effect, since practically all possible final states are unoccupied.
Following these considerations, $g_1$ and $r_2$ can be
obtained as a function of $N$, at a given bias voltage $V$.

The steady state value $\tilde{N}$ of $N$ satisfies charge
conservation in the well, i.e., $g_1(\tilde{N}) = r_2(\tilde{N})$,
and the steady state current is $I =
q g_1(\tilde{N}) = q r_2(\tilde{N})$.

Following Ref. \onlinecite{iannashot97}, it is worth
expanding $g_1(N)$ and $r_2(N)$ around $\tilde{N}$ and
defining the following characteristic times:
\begin{equation}
\frac{1}{\tau_{g}} \equiv - \left. \frac{dg_1}{dN}
                        \right|_{N = \tilde{N}} 
,\hspace{1cm}
\frac{1}{\tau_{r}} \equiv \left. \frac{dr_2}{dN} 
                        \right|_{N = \tilde{N}} \label{taus}
;\end{equation}

Our parameter of choice for studying deviations from full
shot noise is the so-called Fano factor $\gamma$, the ratio of the 
power spectral density of the current noise $S(\omega)$ to the full shot value
$2 q I $. From \onlinecite{iannashot97} we have, in this case,
for $\omega \tau_g \tau_r \ll \tau_g + \tau_r$,
\begin{equation}
\gamma = \frac{S(\omega)}{2 q I} =
         1 - \frac{2 \tau_g \tau_r}{(\tau_g + \tau_r)^2}
.\label{gamma}
\end{equation}
         
From the definition (\ref{taus}), $\tau_{g}$ is positive in
the first region of the I-V characteristic, when Pauli principle and
Coulomb interaction make $g_1$ decreasing with increasing $N$.
On the other hand, in the negative differential resistance
region, the term which varies the most with increasing N is the
longitudinal density of states, which shifts upwards by a factor $q/(C1+C2)$
per electron: since the peak is just below $E_{\rm cbl}$, a
slight shift of the peak sensibly increases the integrand
in (\ref{g1a}), yielding a negative $\tau_{g}$. 
Note that, while from (\ref{gamma}) we see
that noise could also diverge if $\tau_{g} = - \tau_{r}$, 
this cannot physically happen, because the large
deviation of $N$ with respect to $\tilde{N}$ would make
the linearization of $g_1$ and $r_2$ not acceptable.

We now focus on a particular structure, on which we have
performed noise measurements and numerical simulations
following the theory just described. 
Such structure has been  fabricated at
the TASC-INFM laboratory in Trie\-ste
and has the following layer structure:
a Si-doped ($N_d = 1.4 \times 10^{18}$ cm$^{-3}$)
500 nm-thick GaAs buffer layer, an undoped 20~nm-thick
GaAs spacer layer to prevent silicon diffusion into the
barrier, an undoped 12.4~nm-thick AlGaAs first barrier,
an undoped 6.2~nm-thick GaAs quantum well, an undoped
14.1~nm-thick AlGaAs barrier, a 10~nm 
GaAs spacer layer and a Si-doped 500~nm-thick
cap layer. The aluminum mole fraction
in both barriers is $0.36$ and the diameter of the mesa defining
the single device is about $50$~$\mu$m.

The barriers in our samples are
thicker than in most similar resonant-tunneling diodes, 
for the purpose of  reducing the
current and, consequently, to increase the differential resistance,
in order to
obtain the best possible noise match with the measurement amplifiers
(available ultra-low-noise amplifiers offer a good performance, with a very
small noise figure, for a range of resistance values between a few
kiloohms and several megaohms).

We have applied a measurement technique purposedly developed for low-level
current noise measurements, based on the careful evaluation of the
transimpedance between the device under test and the output of the
amplifier. \cite{macupell91}
This procedure allows us to measure noise levels
that are up to 3~dB below that of the available amplifiers with a maximum
error around 10\%.

Our usual approach \cite{macupell91} includes also the
subtraction of the noise due to the amplifier and other
spurious sources, which is evaluated using a substitution
impedance, equivalent to that of the device under test with known
noise behaviour.
For the measurements in the negative differential resistance region,
instead, we have evaluated an upper limit for the noise
contribution from the amplifier in these particular operating
conditions, since it is difficult to sinthesize an appropriate
substitution impedance. From experimental and theoretical
considerations, we have verified that such limit is always below
3~\% of the noise level from the device under test, so that
corrections are not necessary.
In Fig. 2 the measured current and the Fano factor $\gamma$
at the temperature of liquid nitrogen (77~K) are
plotted as a function of the applied voltage (the thicker barrier
is on the anode side).

\begin{figure}
\epsfxsize 8.25cm
\epsffile{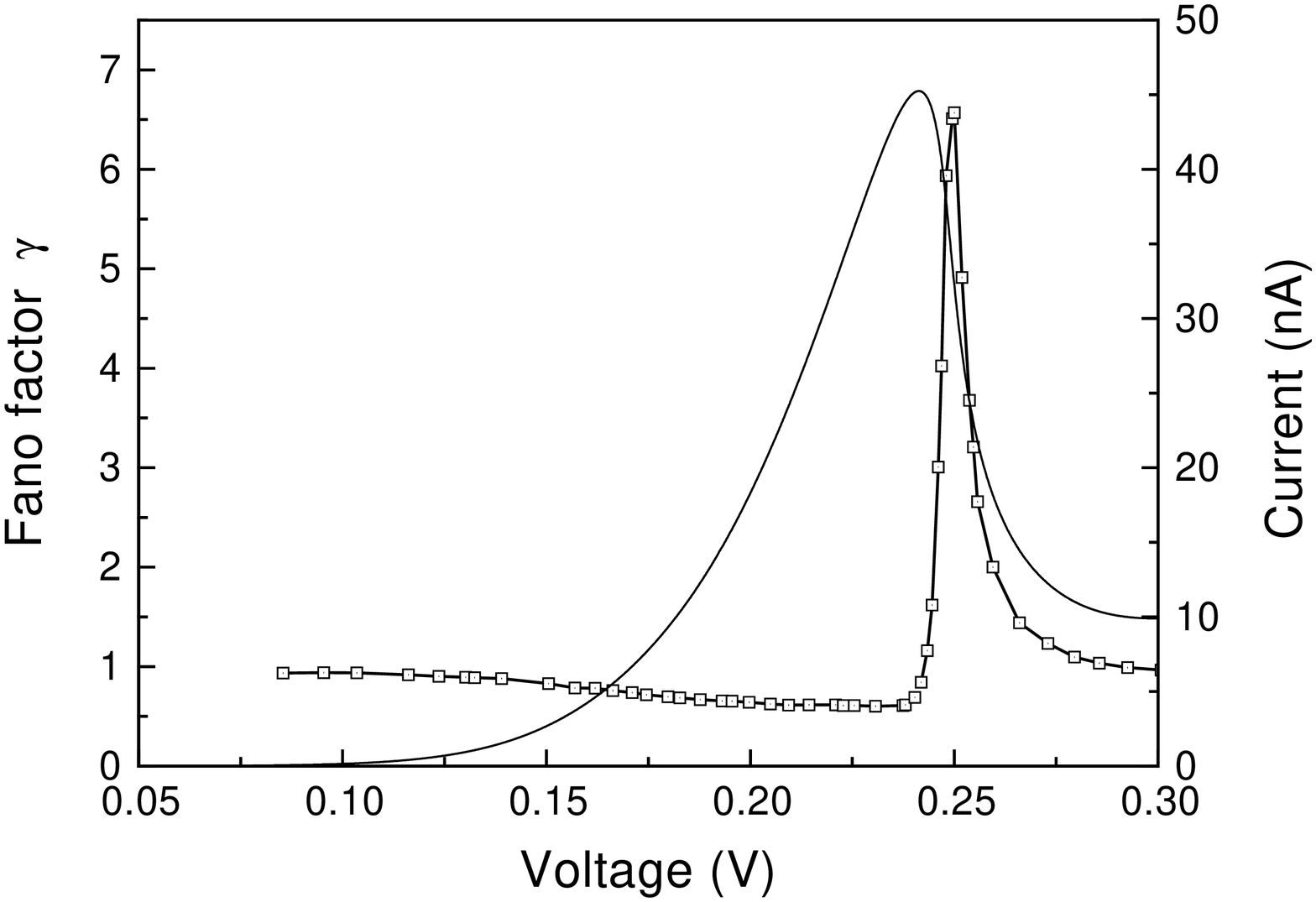}
\vspace{0.2cm}
\caption{Experimental current (solid line) and Fano factor $\gamma$ 
(squares) as a function
of the applied voltage, at the temperature of 77~K. 
The maximum value of $\gamma$ is 6.6, 
while the minimum is close to 0.5}
\end{figure}

It can be noticed that as the voltage increases, the Fano factor
decreases down to about 0.5 (which corresponds to the maximum
theoretical suppression \cite{iannashot97}), at the voltage
corresponding to the current peak is exactly one, then increases
again and reaches a peak of 6.6 at the voltage corresponding
to the lowest modulus of the
negative differential resistance, while, for higher voltages, 
it rapidly approaches one.

In Fig. 3 we show numerical results for the same structure at 77~K
based on the theory discussed before and obtained by considering a 
relaxation length of 15~nm.\cite{ianna_unified95}
As can be seen, there is an
almost quantitative agreement between theory and experiment
( the peak experimental current is 45 ~nA which corresponds
to a current density of 23~A/m$^2$): we ascribe most of the difference
to the tolerance in the nominal device parameters.
All the relevant features of the Fano factor as a function
of the applied voltage are reproduced, and can be easily 
explained in terms of our model.

\begin{figure}
\epsfxsize 8.25cm
\epsffile{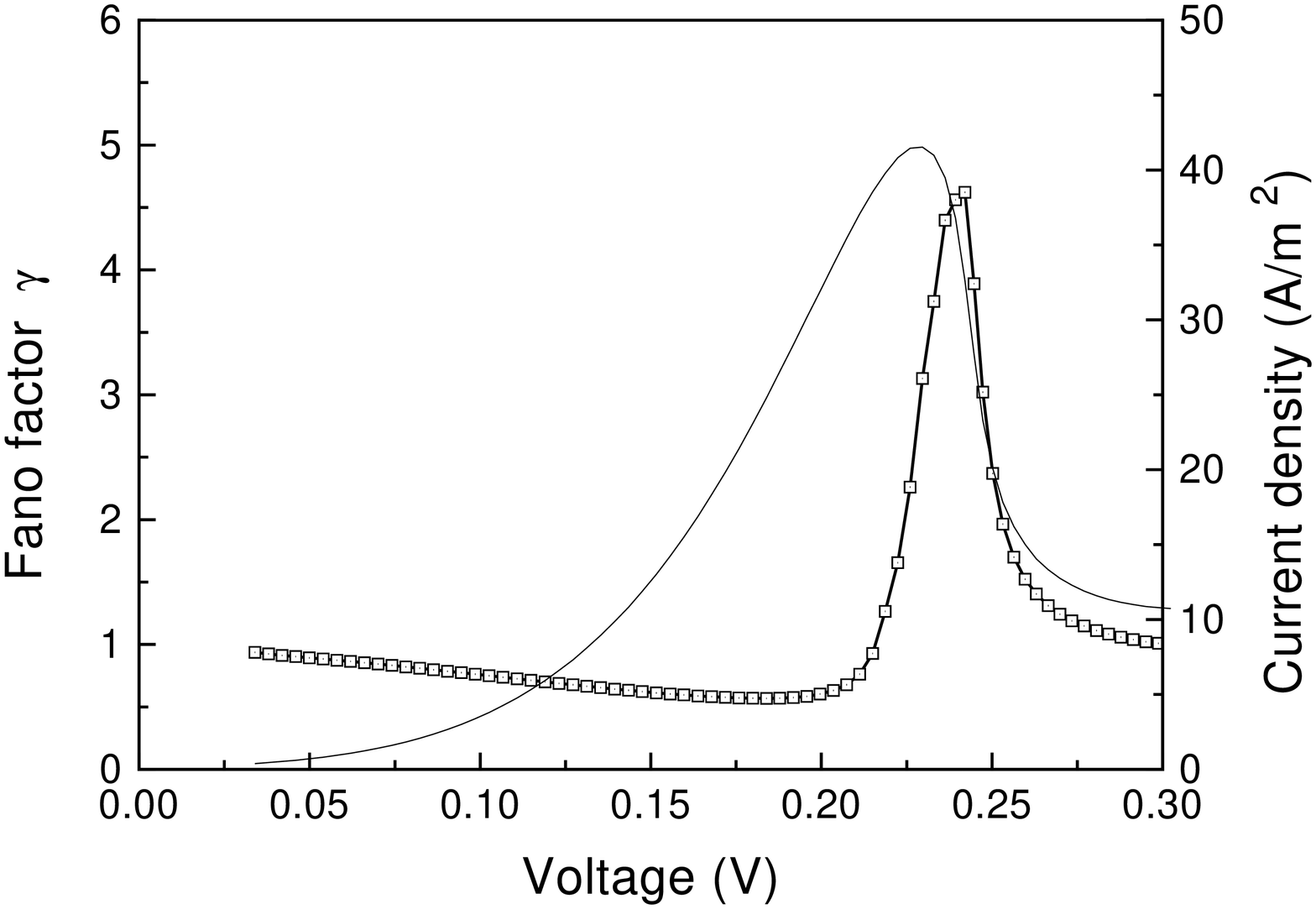}
\vspace{0.2cm}
\caption{Calculated current density (solid line) and Fano factor $\gamma$
(squares) as a function of the applied voltage for the considered structure}
\end{figure}

The fact that $\gamma$ is maximum at the voltage corresponding
to the minimum negative differential resistance $r_d$ of the
device is readily justified once we recognize that $r_d$ is
practically proportional to $\tau_g + \tau_r$ \cite{iannadopo}.
In fact, from Fig. 4 it is clear that $\tau_r$ varies much more smoothly
than $\tau_g$ with the applied voltage, so that,
since $\tau_g$ and $\tau_r$ have opposite sign, 
the modulus of $r_d$ is minimum
when $\tau_g/\tau_r$ approaches $-1$. At this point, we simply notice,
from (\ref{gamma}), that $\gamma$ gets larger as $\tau_g/\tau_r$ 
approaches $-1$, too (and would eventually diverge for $\tau_g = - \tau_r$).

Furthermore, from Figs. 3 and 4, and according to \cite{iannadopo},
we can notice that $r_d$ and $\tau_g$ tend to infinity at about
the same voltage, i.e., the one corresponding to the current peak.
Therefore, according to (\ref{gamma}), $\gamma$ is 1 at the 
current peak bias, as can be verified from both experiments 
and calculations (Figs. 2 and 3, respectively).

\begin{figure}
\epsfxsize 8.25cm
\epsffile{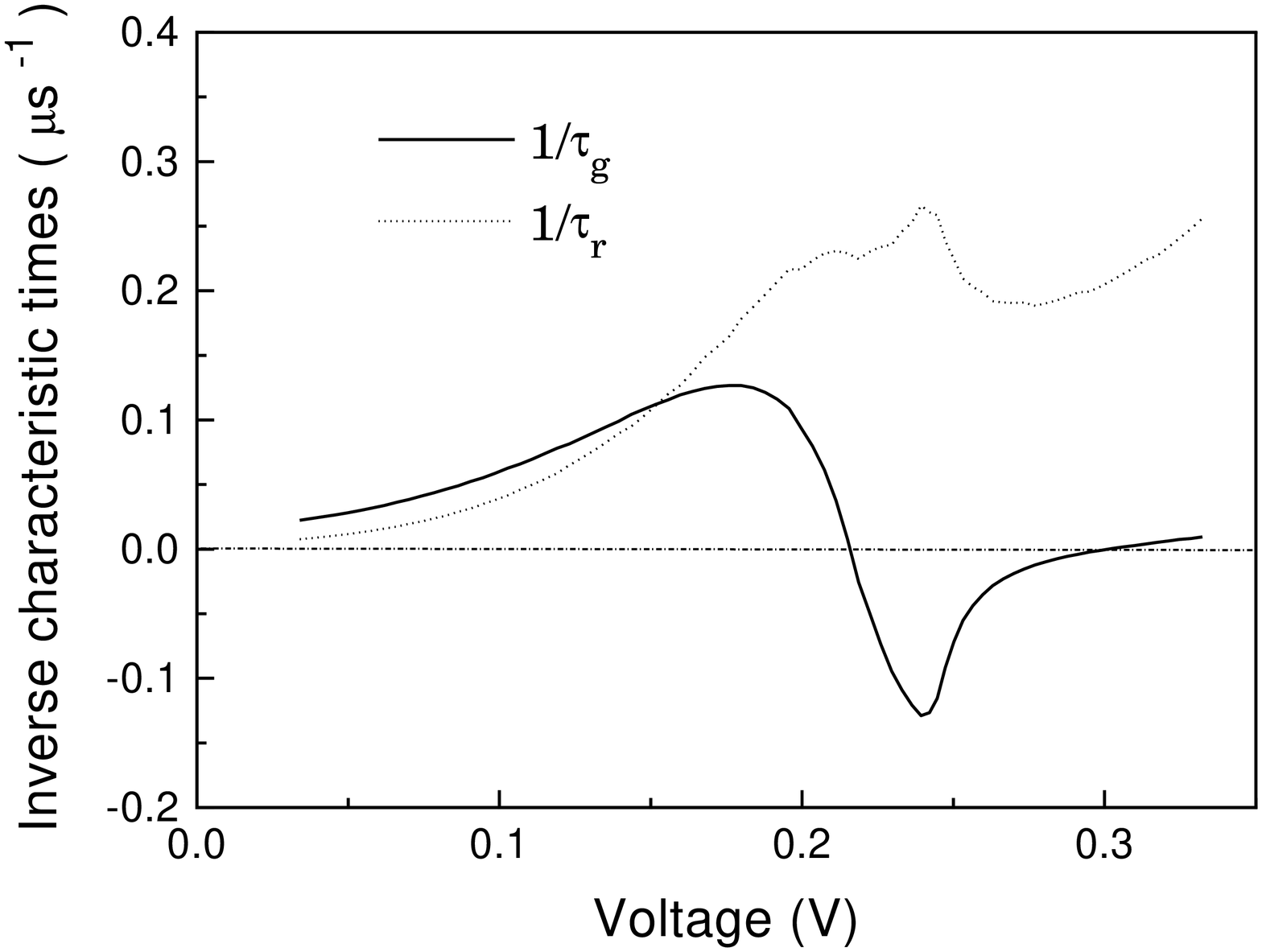}
\vspace{0.2cm}
\caption{Calculated $1/\tau_g$ (solid) and $1/\tau_r$ (dashed) 
as a function of the
applied voltage}
\end{figure}

In conclusion, we have demonstrated experimentally that
Coulomb interaction, enhanced by the shape of the density
of states in the well, can lead to a 
dramatic increase of shot noise in
resonant tunneling diodes biased in the negative differential
resistance region of the I-V characteristic. We have provided a model 
which leads to good numerical agreement with the
experimental data, taking into account all the
relevant physics involved in the phenomenon.


This work has been supported by the Ministry for the University
and Scientific and Technological Research of Italy, 
and by the Italian National Research Council (CNR).


\begin{references}
\bibitem{iannadopo}
We just provide a brief argument:
$$
r_d = \frac{dV}{dI} = 
\frac{1}{q} \left[ 
\frac{d(E_{fl}-E_{fw})}{dI} + \frac{d(E_{fw}-E_{fr})}{dI} 
\right]
;$$
since electrons in the well obey a quasi-equilibrium distribution,
we can assume that the currents through each barrier essentially depend 
on the difference between the quasi-Fermi levels on both sides of the barrier. We therefore
choose to leave $E_{fl}$ and $E_{fr}$ fixed, and let only $E_{fw}$ vary.
We use the fact that $I = q g_1(\tilde{N}) = q r_2(\tilde{N})$ and 
that $E_{fw}$ depends on $N$, to write
$$
r_d \approx \frac{1}{q^2} \left(
\frac{- dE_{fw}/dN}{dg_1/dN} + \frac{dE_{fw}/dN}{dr_2/dN} \right)
=  \frac{dE_{fw}}{dN} \frac{\tau_g +\tau_r}{q^2},
$$
where the last equality comes from (\ref{taus}) and all derivatives
are evaluated for $N = \tilde{N}$.
\end{references}
\end{document}